\documentstyle[12pt]{article}
\begin{document}

\hfill DUKE-CGTP-2000-12

\hfill hep-th/0008150

\vspace{1.5in}

\begin{center}

{\large\bf D-Branes and Scheme Theory}

\vspace{0.5in}


Tom\'as G\'omez \\
Tata Institute for Fundamental Research \\
Homi Bhabha Road, 400 005 Mumbai (India) \\
{\tt tomas@math.tifr.res.in}\\

\hspace{0.1in}  \\

Eric Sharpe \\
Department of Physics \\
Box 90305 \\
Duke University \\
Durham, NC  27708 \\
{\tt ersharpe@cgtp.duke.edu} \\

\hspace{0.1in}  \\

\end{center}

In this highly speculative note we conjecture that it may
be possible to understand some features of coincident D-branes,
such as the appearance of enhanced non-abelian gauge symmetry,
in a purely geometric fashion, using a form of geometry known
as scheme theory.  We give a very brief introduction to some
relevant ideas from scheme theory, and point out how 
these ideas work in special cases.

\begin{flushleft}
August 2000
\end{flushleft}


\newpage

\tableofcontents

\newpage

\section{Introduction}

As is well-known, on $N$ coincident D-branes, $U(1)$ gauge symmetries
are enhanced to $U(N)$ gauge symmetries, and scalars that
formerly described normal motions of the branes become 
$U(N)$ adjoints.  People have often asked what the deep reason for
this behavior is -- what does this tell us about the geometry seen
by D-branes?

One possible answer is that these are indications that
D-branes sense some sort of noncommutative geometry.
However, as this enhancement to nonabelian gauge symmetry occurs
even when $B \equiv 0$, if this is a sign of noncommutative geometry,
it can not quite be noncommutative geometry in the same sense
as \cite{cds,sw}.

In this very short note, we shall speculate that this strange
behavior may, at least in special cases, be an indication that
the geometry seen by D-branes should be understood as a type of
``information-preserving'' geometry known as scheme theory.

Specifically, we begin by giving a very short introduction to
some relevant ideas from scheme theory, including the idea
that coincident subvarieties can be understood as ``non-reduced 
schemes.''  We then point out in a simple example
(originally worked out in \cite{DEL}, for very different reasons)
that branches of the classical moduli space of vacua can be
understood in terms of moduli spaces of rank 1 sheaves on
non-reduced schemes, and that $U(N)$-adjoint-valued scalars
also have a natural understanding. 

Although this may sound somewhat new, in truth part of this is implicit
in the idea that D-branes can be described, at some level,
in terms of coherent sheaves \cite{hm}.  Certainly coherent sheaves
can be used to describe non-reduced subschemes of a variety,
which we shall tentatively identify (in special cases) with
coincident D-branes.  What is new in this paper is that scheme
theory may tie into D-branes at a much deeper level, by
giving a purely geometric means for studying branches of the
classical moduli space of vacua on a D-brane worldvolume,
and by giving a mechanism for understanding how scalars
describing normal motion can become $U(N)$ adjoints. 

We do not claim to describe any particularly new or deep results, mathematical
or physical, in this short note; rather, we shall only give
a few easy calculations and review relevant calculations done
elsewhere.  This note does little more than point out some 
interesting ideas.  We hope to publish more on this topic in the
future.

The recent preprint \cite{viper} contains closely related ideas
to those we shall present here.  We worked on these ideas about two
years ago, but decided to put off trying to write up any part
of it until after a number of thorny technical issues had been
resolved.  However, the publication of \cite{viper} has motivated 
us to rapidly publish this very short overview of our ideas.

We should also pause to mention important differences between
this paper and \cite{viper}.  In \cite{viper}, deformation quantization
was used to generate recognizable D-brane features.
By contrast, deformation quantization never enters this paper.
Instead, we outline how essentially the same features of D-branes
are already present in scheme theory, without resorting to deformation
quantization, albeit in a rather subtle fashion.

A very different approach to understanding the non-abelian structure
of coincident D-branes was also recently published in \cite{hklm}.



\section{Information-preserving geometry}

Scheme theory provides a notion of what we shall call
an ``information-preserving geometry.''
What does that expression mean?

Consider a collision of two points.  When two points collide,
one is left with a single point.  One does not know if a single
point is really some number of coincident points, or if it is 
a single point.  Thus, when thinking about motions of points
(and higher dimensional subvarieties), the usual notions of
geometry lose information -- in the example above, there is
no natural way to tell within the geometry if a single point 
represents a collision of multiple points.
In order to distinguish the two possibilities, one would need
to add additional\footnote{The reader might well object that
this distinction is rather trivial -- one could merely keep track of
multiplicities of points.  However, there is additional information
that scheme theory also encodes, such as the relative tangent
directions of collisions.  We shall -- briefly -- return to this
point later.} information.

By contrast, in scheme theory such information is automatically
encoded in the geometry.  In scheme theory not all points are identical -- 
whether a point represents a collision of several other points
is automatically encoded within the geometry itself.
Thus, scheme theory is an example of an ``information-preserving''
geometry, and is a natural candidate for a purely geometrical
means of distinguishing single D-branes from multiple coincident
D-branes.

\section{The skinny on fat points}

How does scheme theory encode such additional information?
The general idea is to work with the ring of functions\footnote{
In algebraic geometry, algebraic functions.} on a space, rather
than the space itself.  By knowing the functions, one can learn about
the space.

For example, instead of talking about the space ${\bf C}^2$,
one works with the ring of holomorphic functions ${\bf C}[x,y]$.
A single point at the origin in ${\bf C}^2$ is described by the ring
${\bf C}[x,y]/(x,y)$, where $(x,y)$ denotes the ideal generated
by $x$ and $y$.  Coincident points can be described in
a number of different ways, such as ${\bf C}[x,y]/(x^2,y)$
or ${\bf C}[x,y]/(x^2,xy,y^2)$ (depending upon the number of
coincident points, among other things).  The rings describing coincident
points differ from the ring describing a single point, so we see
explicitly how scheme theory distinguishes coincident points
from single, ordinary points.  (Such points are known as
``fat points,'' because they carry more information than a single
point ordinarily would.)  In fact, there is additional information
here.  The multiple ways of describing two coincident points
encodes the relative directions from which they collided -- so
scheme theory can not only tell whether a single topological point
is secretly the collision of several points, but also contains
information about the relative directions from which they came. 

A full introduction to scheme theory is far beyond the intended
scope of this paper.  For the interested reader, the standard
basic reference is \cite{hartshorne}; significantly more readable
accounts, at varying levels of difficulty, include 
\cite{chandler2,eisenbudharris,shaf2,mumfordred}.
In the interests of brevity, our discussion will rapidly become
significantly more technical; however the interested reader
should hopefully still be able to follow our presentation by
frequently consulting the aforementioned references.

A scheme describing multiple coincident varieties is an example
of a ``non-reduced scheme,'' and is an obvious candidate
for a geometric description of coincident D-branes.
This is precisely the conjecture we shall pursue -- that one can
understand many classical features of coincident D-branes in terms
of non-reduced schemes.  (Non-reduced schemes are more general
than merely describing coincident subvarieties, but we shall not
need the more general case here.)

As we are interested in how the $U(1)$ gauge symmetry on the worldvolume
of a single D-brane becomes enhanced to a $U(N)$ gauge symmetry
on $N$ coincident D-branes, one of the first things we should
study is sheaf theory on non-reduced schemes.

With that question in mind,
what sorts of sheaves can appear on a non-reduced scheme?
To be specific, let $A$ be a ring and $I$ a prime ideal in $A$.
Define a non-reduced scheme
$R = \mbox{Spec } A/I^n$, and the corresponding reduced scheme
$C = \mbox{Spec } A/I$.  In this description, $R$ describes
$n$ coincident copies of $C$.

First, any coherent sheaf on $C$ defines a coherent sheaf on $R$
also.  Specifically, it is straightforward to check that
any $(A/I)$-module $M$ is also an $(A/I^n)$-module, hence any coherent
sheaf on $C$ also defines a coherent sheaf on $R$.
However, non-reduced schemes are subtle, and many statements that one
would now be tempted to make are false.  For example, although
a bundle on $C$ defines a coherent sheaf on $R$,
strictly speaking a bundle on $C$ does not define a bundle on $R$.
For example, a rank 1 bundle on $C$ is described by the module
$A/I$, but a rank 1 bundle on $R$ would be described by the module
$A/I^n$, which is distinct.

Another set of examples of coherent sheaves on $R$ are furnished
by ideal sheaves.  For example, take $A = k[x,y]$ for some field
$k$, $I = (x) \subset A$, and define the ideal $J = (x,y)$
(i.e., the ideal generated by $x$ and $y$).
The module $J/I^2 \subset A/I^2$ defines an ideal sheaf on $R$.
It is interesting to note that the restriction of such an ideal
sheaf to $C$ generically has rank 1, but also has a torsion component.
Specifically, the restriction of this ideal sheaf to $C$ is defined
by the module
\begin{displaymath}
(J/I^2) \otimes_{(A/I^2)} (A/I)
\end{displaymath}
This module has two generators, namely $(x + I^2, 1+I)$
and $(y+I^2, 1+I)$, however the generator
$(x+I^2,1+I)$ is annihilated by all elements of $A/I$ other
than the identity.
Thus, this module defines a sheaf on $C$ of the form
${\cal L} \oplus {\cal T}$, where ${\cal L}$ is
a locally free rank 1 sheaf (corresponding to generator
$(y+I^2,1+I)$) and ${\cal T}$ is a torsion sheaf with
support at $x=y=0$ (corresponding to generator $(x+I^2,1+I)$).

Now, it turns out that if $R$ corresponds to $r$ coincident
copies of $C$, then rank $r$ bundles on $C$ often appear
in the same families as rank 1 sheaves on $R$.
(Note that for different sheaves to appear in the same flat 
family means they have
the same Hilbert polynomial (with respect to some fixed projective embedding),
and so appear on the same Hilbert scheme.)
For example\footnote{We would like to thank R.~Donagi for
pointing out this example to us.}, suppose $A = k[x,y]$ and $I = (x)$, as above.
Define a one-parameter family of $(A/I^2)$-modules $M_t$
as follows.  Let $M$ be a freely-generated $(A/I)$-module
with two generators, say $u$ and $v$. 
Construct a family $M_t$ ($t \in k$) of $(A/I^2)$-modules 
by taking the underlying
abelian group to be the same as that for $M$, leaving the $y$ action
on $M$ invariant, and changing $x$ to act on the generators $u$, $v$ as
\begin{eqnarray*}
x: \: u & \longrightarrow & t v \\
x: \: v & \longrightarrow & 0
\end{eqnarray*}
When $t = 0$, we see that $x$ annihilates both generators,
and in fact $M_0 \cong (A/I) \oplus (A/I)$.
When $t \neq 0$, $M_t \cong A/I^2$.

Now that we have set up some basic technology, we shall
describe a simple example.

\section{A simple example}

In \cite{bsv} it was claimed that $r$ D-branes wrapped on a curve $C$
in a K3 of genus $g_C$ were equivalent to a single D-brane
wrapped on a curve of genus $g_R = r^2(g_C - 1) + 1$, lying in
the linear system $| r C |$.  Upon deeper reflection, however,
the reader may wonder about this.  If the twisted bundle endomorphism
$\phi$ is nonzero, then this statement sounds reasonable -- 
the eigenvalues of $\phi$ over any point of $C$ define a cover
of $C$ inside the total space of the canonical bundle on $C$,
and it seems reasonable to think of such $\phi$ as implying that
the D-brane is wrapped on a cover of $C$.

But what about the degenerate limit in which the cover collapses down
to $C$ itself?  In this limit, the cover becomes a non-reduced
scheme $R$, of arithmetic genus $g_R = r^2(g_C - 1) + 1$
(in agreement with the cover).  This situation was discussed
in \cite{DEL}, in the special case $r = 2$. 

We claimed in the introduction that, just as a single D-brane
has a $U(1)$ gauge field, features of $r$ coincident D-branes
can be recovered from Hilbert schemes of rank 1 sheaves on
a corresponding non-reduced scheme.  In \cite[section 3]{DEL},
the Hilbert scheme of rank 1 sheaves on a non-reduced scheme
$R = 2C$ is discussed.  This Hilbert scheme has several
disjoint components, whose closures intersect.  One component
consists of rank 2 bundles\footnote{The reader may be surprised to hear that
a Hilbert scheme of rank 1 sheaves can include rank 2 sheaves, but recall
that we worked out an example of this phenomenon at the end of
the last section.} on $C$, but in addition, there are
other components as well.

The physical interpretation of the first component is clear --
we were expecting to find rank 2 bundles, after all.
But what is the physical interpretation of the other components?

In order to understand the other components, take a moment
to consider the classical moduli space of a two-dimensional
gauge-theory with a single scalar $\phi$ transforming in the adjoint of
$U(2)$.  In particular, suppose $\phi$ is nilpotent, e.g.,
\begin{displaymath}
\phi \: = \: \left( \begin{array}{cc}
                    0 & 1 \\
                    0 & 0 
                    \end{array} \right)
\end{displaymath}
Such a $\phi$ is not gauge-equivalent to the zero matrix,
yet all gauge-invariant observables, such as $\mbox{Tr } \phi$,
$\mbox{Tr } \phi^2$, and $\mbox{det } \phi$ vanish.
So, if we want to study the classical moduli space of such a
gauge theory, we need to be careful to find all possible Higgs branches.
In particular, the ``extra'' components of a Hilbert scheme
of rank 1 sheaves on $R = 2C$ correspond precisely to such
nilpotent Higgs branches.

So far we have described components of Hilbert schemes of rank 1
sheaves on a non-reduced scheme $R = 2C$, and have given intuitive
explanations of what these components should correspond to morally.
However, a stronger statement can be made.

In \cite[section 1]{DEL}, a precise description is given for how
to explicitly deform Hilbert schemes of rank 1 sheaves on a
non-reduced scheme $R = 2C$ to solutions of Hitchin's equations
for which the twisted-bundle-endomorphism $\phi$ is nilpotent
(the ``nilpotent cone'').
In effect, \cite[section 1]{DEL} describes how to recover
scalars that look like $N \times N$ matrices from non-reduced schemes.
(Admittedly, however, such a description is implicit, not explicit,
in their work.)  In particular, our ``intuitive explanations''
for the components of the Hilbert scheme of rank 1 sheaves on
$R$ have a firm footing.

\section{Nonlocality}

One of the oft-quoted reasons for the popularity of
noncommutative geometry is a certain notion of nonlocality.
There is also a notion of nonlocality in scheme theory.

More precisely, by ``nonlocality'' we are referring to the
fact that in scheme theory, there is typically
no notion of a ``small'' open set -- all open sets are very large.
The essential difficulty
is that closed sets are specified as vanishing
loci of holomorphic functions -- so closed sets are complex codimension
one or higher.  Therefore, all open sets see most of the space.

Put another way, the topologies that are most natural here are not
the same topologies that one typically learns about in the
context of differential geometry, for example.

This fact leads to a number of nontrivial technical complications
when working in scheme theory.  For example, since there are no
``small'' open sets, there is no such thing as a tubular neighborhood.
Workarounds exist for these difficulties -- instead of trying to
use tubular neighborhoods to study locally how subvarieties
sit inside an ambient space, one instead uses deformation arguments
(in this case, deformation to the normal cone).

We do not have any strong statements to make about this notion
of nonlocality, its physical interpretation in general,
or its relationship to the nonlocality bandied about
in noncommutative geometry specifically.
We are merely pointing
that ideas of nonlocality, at least loosely analogous
to those in noncommutative geometry, also exist in scheme theory.

\section{Too much information?}

One significant problem with any attempt to describe
coincident D-branes in terms of non-reduced schemes is that
non-reduced schemes appear to encode more information than D-branes
do physically.  More precisely, when subvarieties of some space
collide, the resulting non-reduced scheme not only encodes the 
number of coincident subvarieties, but also the relative directions
from which they came.  (For example, it is this extra information
that distinguishes a Hilbert scheme of $N$ points from
a symmetric product.)  Unfortunately, it is difficult to see
how this extra information could appear physically.

The answer to this puzzle might be that scheme theory is only
relevant for certain $B$ field backgrounds, analogous 
(perhaps even equivalent) to the occurrence of noncommutative geometry
\cite{cds,sw}.

As a simple example, it was pointed out in \cite{ns} that this precise
difficulty is solved for fat points by turning on a $B$ field.
More precisely, in \cite{ns} it was argued that in the presence
of a constant $B$ field background,
torsion-free sheaves on ${\bf C}^2$ appear physically
(as ``noncommutative $U(1)$ instantons'').
(The $B$ field appearing in \cite{ns} was of type (1,1), and so
only generated noncommutativity between holonomorphic and
nonholomorphic sectors.  As emphasized to us by N.~Nekrasov,
if one restricts to holomorphic coordinates, the resulting
ring is commutative, and so the noncommutative $U(1)$
instantons are truly the same thing as torsion-free sheaves.)
Now, torsion-free sheaves encode the same information
as fat points -- the information about relative collision directions
that confused us two paragraphs above appears physically as moduli
of coincident noncommutative $U(1)$ instantons.

From the discussion in \cite{ns}, we are led to suspect
that scheme theory will only be physically relevant for
certain
nonzero $B$ field backgrounds, as such backgrounds naively seem
to enable us to encode the information naturally present in
scheme theory.

\section{Conclusions and bolder conjectures}

In this short note we have conjectured that scheme theory
may be relevant for studying D-branes, and given some limited
evidence for this proposal.

If indeed scheme theory is relevant for describing D-branes
classically, one is naturally led to several further conjectures:
\begin{enumerate}
\item There is, at least naively, a natural candidate within scheme
theory to understand large N limits.  Namely, it seems very natural
to try to describe large N limits in terms of completions,
or formal schemes \cite[section II.9]{hartshorne}.
(This idea was also pointed out in \cite{viper}.)

\item 
One might be able to use scheme theory to shed new light
on wrapped D-branes.
For example, one can build schemes that
are morally K3 surfaces (in that their dualizing sheaves are
trivial and $H^1$ of their structure sheaf vanishes),
but whose underlying topological space is merely some
rational normal scroll \cite{bayereisen,k3carpet1}.
(Such schemes are known in the literature as K3 carpets.)
Perhaps this is some scheme-theoretic way to think about
the topological twisting on wrapped D-branes pointed out in
\cite{bsv}.

\item It might be possible to explicitly see small instantons
splitting off and becoming D-branes (described as torsion sheaves),
as different components of
some Hilbert scheme.

\item If the nonabelian gauge symmetry on D-branes has a scheme-theoretic
understanding, then perhaps the $\mbox{Spin}(32)/{\bf Z}_2$
gauge symmetry of type I strings also has a scheme-theoretic
understanding, from viewing type I strings as D9-branes.
It is difficult to see how such a gauge symmetry could arise
in complex algebraic geometry; perhaps this arises from analogous
ideas in real algebraic geometry.

\item One of the reasons mathematicians are fond of scheme theory
is that it gives a unified framework in which to study both
algebraic geometry and number theory.  Perhaps some sort of
scheme-theoretic approach to D-branes could be used to shed
light on the number-theoretic ideas in \cite{ana1,ana2}. 
\end{enumerate}

We would again like to emphasize that we do not presently feel the ideas
presented in this short note to be more than enticing conjectures.
We hope to report on further developments along these lines at
some point in the future.

\section{Acknowledgements}

We would like to thank P.~Aspinwall, R.~Donagi, D.~Morrison,
and R.~Plesser for useful conversations.

This research was partially supported by National Science Foundation
grant number DMS-0074072.

\end{document}